\documentclass[showpacs,prb,preprintnumbers,amsmath,amssymb]{revtex4}
\usepackage{epsfig}
\usepackage{graphicx}
\usepackage{dcolumn}
\usepackage{bm}
\newcommand{\vnabla}{{\mbox{\boldmath$\nabla$}}}

\newcommand{\vR}{{\mbox{\boldmath$R$}}}

\newcommand{\vA}{{\mbox{\boldmath$A$}}}

\newcommand{\vk}{{\mbox{\boldmath$k$}}}
\newcommand{\vv}{{\mbox{\boldmath$v$}}}

\newcommand{\vB}{{\mbox{\boldmath$B$}}}

\newcommand{\vsk}{{\small \mbox{\boldmath$k$}}}

\newcommand{\vg}{\mbox{\boldmath$g$}}
\newcommand{\vq}{\mbox{\boldmath$q$}}

\newcommand{\vp}{\mbox{\boldmath$p$}}
\newcommand{\hvk}{\hat{\mbox{\boldmath$k$}}}
\newcommand{\hvg}{\hat{\mbox{\boldmath$g$}}}

\newcommand{\vsig}{\mbox{\boldmath$\sigma$}}

\newcommand{\vH}{\mbox{\boldmath$H$}}
\newcommand{\vS}{\mbox{\boldmath$S$}}
\begin{document}

\title{Stability of magnetic-field induced helical phase in Rashba superconductors}
\author{D.F. Agterberg and R.P. Kaur}
\address{Department of Physics, University of Wisconsin-Milwaukee, Milwaukee, WI 53211}

\begin{abstract}
Due to the lack of both parity and time reversal symmetries, the
Rashba superconductors CePt$_3$Si, CeRhSi$_3$, and CeIrSi$_3$, in
the presence of a magnetic field, are unstable to helical (single
plane-wave) order. We develop a microscopic theory for such
superconductors and examine the stability of this helical phase.
We show that the helical phase typically occupies
most of the magnetic field-temperature phase diagram. However, we
also find that this phase is sometimes unstable to a multiple-{\it
q} phase, in which both the magnitude and the phase of the order
parameter are spatially varying. We find the position of this
helical to multiple-{\it q} phase transition. We further examine
the density of states and identify features unique to the helical
phase.
\end{abstract}
\pacs{74.20-z, 71.18.+y} \maketitle

\noindent

\section{Introduction}

Recently, a new class of heavy fermion
superconductors have been discovered that break parity symmetry
\cite{bau04,kim05,sug06}. The broken parity symmetry implies that
the three materials CePt$_3$Si, CeRhSi$_3$, and CeIrSi$_3$ all
allow a Rashba spin-orbit coupling. The energy scale of this
coupling is much larger than the superconducting energy scale
\cite{sam04,kim01}. This has many non-trivial implications on the
resulting superconducting state
\cite{ede89,gor01,agt03,bar02,yip02,dim03,fri04,min04,sax04,ser04,sam04-2,kau05,fuj05}.
Of particular relevance to the work presented here, are the
unusually high upper critical fields in these three Rashba
superconductors. These fields substantially exceed the usual Pauli
limiting field. Consequently, the Zeeman interaction must play an
important role in the the physics of the superconducting state.
Some progress has been made in identifying the ground states of
Rashba superconductors in Zeeman fields. In particular,
microscopic arguments have been presented that indicate that a
superconducting 'stripe' phase appears in two-dimensional (2D)
$s$-wave Rashba superconductors when a Zeeman field is applied in
the plane \cite{bar02,dim03}. This phase resembles the Zeeman
field induced phase for conventional superconductors found
theoretically by Fulde, Ferrell, Larkin, and Ovchinnikov (FFLO)
\cite{ff,lo} which has recently been observed in CeCoIn$_5$
\cite{bia03,rad03}. In the FFLO phase, the superconducting order
vanishes periodically along a line, taking the simple form
$\Delta(\vR)=\Delta_0\cos(\vq\cdot \vR)$ near the upper critical
field. As the field is reduced, more Fourier components appear in
the order parameter of this phase and, at low enough fields, a
true stripe like order develops \cite{lo}. We will call this phase
(and its appropriate generalization to Rashba superconductors) the
multiple-{\it q} phase. However, phenomenological arguments for
Rashba superconductors indicate that a helical phase is the stable
ground state near $T_c$, once a magnetic field is applied
\cite{agt03,kau05,sam04-2}. In this phase, the order parameter
takes the form $\Delta(\vR)=\Delta_0e^{i\vq\cdot\vR}$. The gap
magnitude is spatially homogeneous and therefore the helical phase
exhibits quite different physical properties than that of the
multiple-{\it q} phase. To address the physical properties of
Rashba superconductors in magnetic fields, it is important to
understand which of these phases are stabilized.

Vortices will also play an important role in understanding the
physics of Rashba superconductors in magnetic fields. This work
focuses on the role of a Zeeman field on Rashba superconductors.
Previous experience has shown that the physics associated with the
Zeeman field persists when vortices are present. In particular,
the Zeeman field induced FFLO phase coexists with vortices in the
heavy fermion superconductor CeCoIn$_5$ \cite{bia03,rad03}.
Furthermore, the helical phase discussed above has been shown to
coexist with vortices in Rashba superconductors near the upper
critical field \cite{kau05,sam04-2}.

\begin{figure}
\epsfxsize=2.0 in \center{\epsfbox{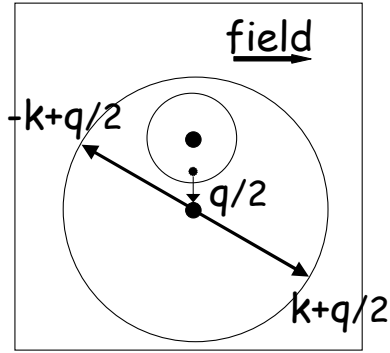}} \caption{A
magnetic field directed as shown shifts the center of the two
Fermi surfaces by $\pm\vq/2$. The smaller dot represent the point
$(0,0)$ (center of Fermi surfaces without field) and the two
larger dots represent the points $(0,-\vq/2)$ and $(0,\vq/2)$
(these are the centers of the new Fermi surfaces). To gain
condensation energy, pairing occurs between states of $\vk+\vq/2$
and $-\vk+\vq/2$, leading to a gap function that has a spatial
variation $\Delta(\vR)=\Delta_0\exp(i\vq\cdot\vR)$.} \label{fig1}
\end{figure}\vglue 0.5 cm

To understand the microscopic origin of the helical and
multiple-{\it q} phases, it is useful to consider the
quasiparticle states when inversion symmetry is broken and a
Zeeman field is applied. Broken inversion symmetry allows an
antisymmetric spin-orbit coupling:
$\alpha\vg(\vsk)\cdot\vS_{\vsk}$ for a quasiparticle with spin
$\vS_{\vsk}$ and momentum $\vk$. Note that
$\vg_{\vsk}=-\vg_{-\vsk}$ due to time reversal invariance. For a
Rashba interaction, $\vg_{\vsk}=(k_y,-k_x,0)/k_F$. The addition of
the Zeeman field leads to the additional coupling
$\mu_B\vH\cdot\vS_{\vsk}$ and the quasiparticle energy eigenvalues
become $E_{\vsk,\pm} = \xi_{\vsk} \pm |\alpha\vg_{\vsk}+\mu_B\vH|$
where $\xi_{\vsk}$ is the quasiparticle energy when
$\alpha=\mu_BH=0$. We are interested in Zeeman fields that are
comparable to the gap energy scale and therefore consider the
limit $\alpha>>\mu_BH$. For a cylindrical Fermi surface, with a
Rashba interaction, applying a field along the $\hat{x}$
direction, we find $E_{\vsk,\pm}=k_x^2/2m+ (k_y\pm
q/2)^2/2m\pm|\alpha\vsk|/k_F$ where $q=2m\mu_BH/|\vk|$. The key
point is that the Fermi surfaces remain circular and the {\it
centers of the two Fermi surfaces are shifted along $\hat{y}$ in
opposite directions}.  In this situation, one of the bands can
gain condensation energy by pairing fermions through the new
center of the appropriate Fermi surface. This leads to the helical
phase in which the condensate wavefunction  becomes
$\Delta(\vR)=\Delta_0\exp(i\vq\cdot\vR)$. This situation is
depicted in Fig.~1. However, the gain in condensation energy of
one Fermi surface is accompanied by a corresponding loss in
condensation energy on the other Fermi surface, since the centers
are shifted in opposite directions. This leads to a competition
between the stability of the helical phase and that of the
muliple-{\it q} phase (in which both $\pm\vq$ gap function modes
appear). In this paper, we address the resulting phase diagram by
developing the quasi-classical Eilenberger equations for a Rashba
superconductor in the limit $\alpha>>\mu_B H$. We further examine
the density of states of the helical phase.

\section{Microscopic formulation}

We consider the following Hamiltonian:
\begin{equation}
{\cal H}_0= \sum_{\vsk,s,s'}c^{\dagger}_{\vsk,s} \left
\{\xi_{\vsk} \sigma_0 + [\alpha
  \vg_{\vsk} +\mu_B\vH] \cdot \vsig \right\}_{ss'}
c_{\vsk,s'}+ \frac{1}{2} \sum_{\vsk, \vsk,\vq} V(\vk,\vk')
c^{\dag}_{\vsk+\vq/2,\uparrow}
                     c^{\dag}_{-\vsk+\vq/2,\downarrow} c_{-\vsk'+\vq/2,\downarrow}
                     c_{\vsk'+\vq/2,\uparrow}.
\end{equation}
We also set $ \langle \vg_{\vsk}^2 \rangle = 1 $ where $\langle
\rangle $ represents an  average over the Fermi surface.  We have
restricted ourselves to spin-singlet pairing interactions since
this is sufficient to capture the new physics associated with the
helical phase. In the large $\alpha$ limit, $\alpha>>T_c$, the
pairing problem becomes a real two-band problem in the diagonal
spinor ($\pm$) basis. In this basis, the pairing interaction
becomes (this is after redefining the gap functions and the
anomalous propagators by a $\vsk$ dependent phase factor
\cite{ser04}),
\begin{equation}
V= \frac{1}{2} V(\hvk,\hvk') \left( \begin{array}{cc} 1& -1\\
-1& 1\end{array} \right) \;
 \label{twoband}.
\end{equation}

We work within the quasi-classical approximation and define the usual Green's functions in Nambu space for each band ${\bm \Psi}_{\pm}^\dagger(\vR)=[\psi^\dagger_{\pm}(\vR),\psi_{\pm}(\vR)]$ and define the imaginary time Green's function \begin{equation}
\hat{G}_{\pm}({\bm x}_1,{\bm x}_2;\tau_1-\tau_2)=-\langle T_\tau {\bm \Psi}_{\pm}({\bm x}_1,\tau_1) {\bm \Psi}_{\pm}^\dagger({\bm x}_2,\tau_2)\rangle,
\label{greens-func}
\end{equation}
here the operator $T_\tau$ arranges the field operators in ascending order of the imaginary time $0<\tau<1/T$ and $\bm{\Psi}({\bm x},\tau)=e^{\tau \cal{H}}\bm{\Psi}(\bm x)e^{-\tau\cal{H}}$. We introduce the center-of-mass coordinate, ${\bm R}=({\bm x}_1+{\bm x}_2)/2$ and the relative coordinate, ${\bm r}={\bm x}_1-{\bm x}_2$, and perform the Fourier transformation in the latter according to
\begin{equation}
\hat{G}_{\pm}(\vk,\vR;\omega_n)=\int d{\bm r}\int_0^{1/T} d{\tau} \hat{G}_{\pm}\left({\bm x}_1,{\bm x}_2;\tau\right)
e^{-i({\bm k}\cdot{\bm r}-\omega_n\tau)},
\end{equation}
where $\omega_n=(2n+1)\pi T$ is the fermionic Matsubara frequency. We define
\begin{equation}
\hat{g}_{\pm}(\hvk,\vR,\omega_n)=
\left(\begin{array}{cc}
g_{\pm} & f_{\pm} \\
f^\dagger_{\pm} & -g_{\pm}
\end{array}\right)
\equiv\frac{i}{\pi}\int d\xi\hat{\tau_3}\hat{G}_{\pm}(\vk,\vR,\omega_n),
\end{equation}
where $d\xi$ integrates out the variable perpendicular to the Fermi surface, $\hvk$ is vector on the Fermi surface,  and $\tau_3$ is the $z$-component of the Pauli matrices acting on the particle-hole space. The standard quasi-classical approach \cite{eil68,lar68,ser83}
results in the following Eilenberger equations for this system :
\begin{equation}
[\omega_n\pm i \mu_B\hvg_{\hvk}\cdot \vB +
\vv_{\hvk}\cdot(\vnabla+\frac{2ie}{
c}\vA)]f_{\pm}=\Delta_{\pm}(\hvk,\vR)g_{\pm} \label{eilen1}
\end{equation}
\begin{equation}
[\omega_n\pm i \mu_B\hvg_{\hvk}\cdot \vB -
\vv_{\hvk}\cdot(\vnabla-\frac{2ie}{
c}\vA)]f_{\pm}^{\dagger}=\Delta^*_{\pm}(\hvk,\vR)g_{\pm}
\label{eilen2}
\end{equation}
where $f_{\pm}^{\dagger}f_{\pm}+g_{\pm}^2=1$, $\omega_n=\pi
T(2n+1)$, $\hvk$ denotes $\vsk$ restricted to the Fermi surface,
$\vA$ is the vector potential (included for completeness), and
$\vB=\nabla\times\vA$.  We will neglect $\vA$ and set $\vB=\vH$ in
the following. With the two band pairing interaction of
Eq.~\ref{twoband}, the gap equation is
\begin{equation}
\Delta_i(\hvk,\vR)=-\pi
T\sum_{n,j}N_j<V_{ij}(\hvk,\hvk')f_j(\hvk',\vR,\omega_n)>_{\hvk'}\label{gap}
\end{equation}
where $N_j$ is the density of states on band $j$. We also express
$V(\hvk,\hvk')=-V \varphi_{\Gamma}(\hvk)\varphi^*_{\Gamma}(\hvk')$
where $\Gamma$ labels the 1D gap representation we are interested
in.  The gap function can be written in the product form
$\Delta_{\alpha}(\hvk,\vR)=\tilde{\Delta}_{\alpha}(\vR)\varphi_{\Gamma}(\hvk)$
where Eq.~\ref{twoband} implies that
$\tilde{\Delta}_{+}(\vR)+\tilde{\Delta}_{-}(\vR)=0$, indicating
that the gap magnitudes on the two bands are equal at every
position in space. Henceforth we set
$\Delta(\vR)=-\tilde{\Delta}_+(\vR)=\tilde{\Delta}_-(\vR)$.

The Eilenberger equations can be derived from a Gibbs free-energy
functional \cite{eil68}. We will require this free energy to
compare the energies of the different phases. Once
Eqs.~\ref{eilen1} and \ref{eilen2} are solved for a given
functional form of $\Delta(\vR)$, this free energy functional
becomes
\begin{equation}
\Omega_{\rm SN}= \int d{\bm R} \left[V^{-1}
\left|\Delta(\vR)\right|^2 -\pi T \sum_{n,j}N_j \left\langle
I_j(\hvk,\vR,\omega_n)\right\rangle\right], \label{free-energy}
\end{equation}
with
\begin{equation}
I_j(\hvk,\vR,\omega_n)= \frac{\varphi^*_{\Gamma}(\hat{\bm
k}){\Delta}^*(\bm R)f_j(\hvk,\vR,\omega_n)
+f_j^\dagger(\hvk,\vR,\omega_n){\Delta}({\bm
R})\varphi_{\Gamma}(\hat{\bm
k})}{1+\frac{\omega_n}{|\omega_n|}g_j(\hvk,\vR,\omega_n)}
\label{func-i}
\end{equation}

\section{Analysis of the helical phase}
\subsection{Helical phase solution}

Eqs. \ref{eilen1} and \ref{eilen2} can be solved for $\Delta(\hvk,\vR)=\Delta \varphi_\Gamma(\hvk)e^{i\vq\cdot\vR}$,  the solutions are:
\begin{equation}
g_{\pm}=\frac{\tilde{\omega_n}_{\pm}}{\sqrt{\tilde{\omega_n}_{\pm}^2+|\Delta\varphi_\Gamma(\hvk)|^2}}
\end{equation}
and
\begin{equation}
f_{\pm}=\frac{\Delta\varphi_{\Gamma}(\hvk)e^{i\vq\cdot\vR}}{\sqrt{\tilde{\omega_n}_{\pm}^2+|\Delta\varphi_\Gamma(\vk)|^2}}
\end{equation}
where $\tilde{\omega_n}_\pm=\omega_n\pm
i\mu_B\hvg\cdot\vH+i\vv\cdot\vq$. The free energy
Eq.~\ref{free-energy} is then minimized with respect to $\Delta$
and $\vq$ to find the $H$-$T$ phase diagram. This phase diagram
depends strongly upon field orientation, dimensionality, and the
relative values of $N_-$ and $N_+$.  It depends weakly upon the
pairing symmetry. For field parallel to $\hat{z}$, the stable
state is given by $\vq=0$ and the physics is independent of the
Zeeman field \cite{kau05}. In the following, we consider only
magnetic fields applied perpendicular to $\hat{z}$. When $N_+\ne
N_-$, then $\vq\ne 0$ whenever $\vH\ne 0$. Some analytical results
can be found for the upper critical field. For a 2D cylindrical
Fermi surface, independent of pairing symmetry,  the upper
critical field diverges as $T\rightarrow 0$ and $v_F\vq=\mu_B
\hat{z}\times\vH$ at $H_{c2}$ for $N _+>N_-$ (a related result has
been found previously ~\cite{bar02,dim03}). We have further found
that if both attractive spin-triplet and attractive isotropic
spin-singlet pairing interactions are included, then the
divergence of $H_{c2}$ occurs occurs at $T>0$ for a 2D cylindrical
Fermi surface. In particular, if $N_+=N_-$, then this divergence
occurs when $T=\sqrt{T_sT_p}$ where $T_s$ ($T_p$)is the usual
$T_c$ for isotropic $s$-wave ($p$-wave) pairing when $\alpha=0$.
For an isotropic $s$-wave superconductor, with a 3D spherical
Fermi surface at $T=0$, $H_{c2}=2 H_P e^{1+\pi|\delta N|/2}$ where
$\delta N=(N_+-N_-)/(N_++N_-)$, and
$H_P=\frac{\Delta_0}{\sqrt{2}\mu_B}$ is the usual Pauli
paramagnetic field , and $\Delta_0=1.76 T_c$. Generically, the
appearance of the helical phase enhances the upper critical field
well beyond the Pauli limiting field as is observed in CePt$_3$Si
\cite{yas04,kau05}, CeIrSi$_3$ \cite{sug06}, and CeRhSi$_3$
\cite{kim05}.

\subsection{Stability of the helical phase}
As explained in the Introduction, there are physical reasons to suggest that the
helical phase is not always stable.  To examine this possibility,
we set $\Delta(\vR)=\Delta_q e^{i\vq\cdot
\vR}+\Delta_{p}e^{i\vp\cdot\vR}+\Delta_{2q-p}e^{i(2\vq-\vp)\cdot
\vR}$ where $\vq$ is the Fourier component that optimizes the
helical phase free energy. The other two Fourier modes represent
the instability of the helical phase to a phase we name the
multiple-{\it q} phase. With this solution for $\Delta(\vR)$, we solve
the Eqs. \ref{eilen1} and \ref{eilen2} perturbatively, and expand
the free energy in Eq.~\ref{free-energy} to second order in these
two modes.

To carry out this procedure, we re-write the Eilenberger equations in terms of Fourier components. In particular, we set $\Delta(\vR)=\sum_{q}e^{i\vq\cdot\vR}\Delta_q$, $g(\vR)=\sum_qe^{i\vq\cdot\vR}g_{q}$, and
$f(\vR)=\sum_{q}e^{i\vq\cdot\vR}f{q}$ (we have suppressed the $\alpha$, $\hvk$ and $\omega_n$ labels for notational simplicity). The Eilenberger equations become
\begin{equation}
\omega_q f_{q}=\sum_{p} \Delta_{p} g_{q-p}
\end{equation}
\begin{equation}
\omega_{-q}f^{\dagger}_{q}=\sum_{p} \Delta^*_{p} g_{q+p}
\end{equation}
\begin{equation}
\sum_{q} [g_{q+p}g_{-q}+f_{q+p}f^{\dagger}_{-q}]=\delta_{p,0}
\end{equation}
where we have defined $\omega_q=\omega_n\pm i \mu_B\hvg_{\hvk}\cdot \vB +i\vv_{\hvk}\cdot\vq$ (the $\pm$ refer to the two different bands). We wish to solve these equations as a perturbation about the helical phase $\Delta_Q\ne 0$ (we use gauge invariance to choose this real), keeping terms in the free energy up to second order in $\Delta_{q}$ and $\Delta_{2Q-q}$. To carry this out, we require $f_q,f^{\dagger}_{-q},g_{q-Q}$, and $g_{Q-q}$ to first order in the perturbation and $g_0$ and $f_{Q}+f^{\dagger}_{-Q}$ to second order in the perturbation. We label the second order corrections to $g_0,f_{Q}+f^{\dagger}_{-Q}$ as $\tilde{g}_0,\tilde{f}_{Q}+\tilde{f}^{\dagger}_{-Q}$ and keep the labels $g_0,f_{Q},f^{\dagger}_{-Q}$ for the zeroth order perturbation. A lengthy calculation gives
\begin{equation}
f_{q}=\frac{g_0\left[(2\omega_{Q}\omega_{2Q-q}+\Delta_{Q}^2)\Delta_{q}-\Delta_{Q}^2\Delta_{2Q-q}^*\right]}
{2\omega_{Q}\omega_{q}\omega_{2Q-q}+\Delta_{Q}^2(\omega_{q}+\omega_{2Q-q})},
\end{equation}
\begin{equation}
g_{q-Q}=-\frac{g_0\Delta_Q\left[\omega_{2Q-q}\Delta_q+\omega_q\Delta_{2Q-q}^*\right]}
{2\omega_{q}\omega_{2Q-q}\omega_Q+\Delta_{Q}^2(\omega_{q}+\omega_{2Q-q})},
\end{equation}
\begin{equation}
\tilde{f}_Q+\tilde{f}^{\dagger}_{-Q}=\frac{g_0g_{q-Q}(\Delta_q+\Delta_{2Q-q}^*)+g_0g_{q-Q}(\Delta_q^*+\Delta_{2Q-q})-2\Delta_Qg_{q-Q}g_{Q-q}-
f^{\dagger}_{-q}f_q\Delta_Q-f^{\dagger}_{-2Q+q}f_{2Q-q}\Delta_Q}{g_0\omega_Q+\Delta_Qf_Q},
\end{equation}
and
\begin{equation}
\tilde{g}_0=-\frac{2g_{q-Q}g_{Q-q}+f_Q(\tilde{f}_Q+\tilde{f}^{\dagger}_{-Q})+f^{\dagger}_{-q}f_q+f^{\dagger}_{-2Q+q}f_{2Q-q}}{2g_0}.
\end{equation}
Using these expressions in Eq.~\ref{free-energy} leads to a free energy of the form
\begin{equation}
\alpha_p|\Delta_p|^2+\alpha_{2q-p}|\Delta_{2q-p}|^2+\alpha_m\Delta_p\Delta_{2q-p}+\alpha_m^*\Delta_p^*\Delta_{2q-p}^*
\end{equation}
The helical phase is unstable when the above free energy becomes
negative for any choice of $\Delta_p$ or $\Delta_{2p-q}$. We find
numerically that the instability occurs for $\vp=-\vq$. Fig.~2
shows the phase diagram found for $N_+=N_-$ ($\delta N=0$) for
isotropic pairing with a 3D spherical Fermi surface. In this case,
the helical phase occupies only a small region of the phase
diagram. We find that the transition from the uniform phase into
the helical phase is first order for $T/T_c<0.39$. However, for
$T<0.36$, we find that the helical phase is itself unstable to the
multiple-{\it q} phase. Fig.~2 also shows the phase diagram found
when $\delta N=0.05$. Contrary to $\delta N=0$, the helical phase
occupies almost the entire phase diagram, with the multiple-{\it
q} phase existing in a region at low temperature and moderate
fields. The region occupied by the stripe phase decreases as
$\delta N$ increases. Once $\delta N>0.25$, we find that the
multiple-{\it q} phase ceases to exist. The suppression of the
multiple-{\it q} phase due to an increasing $\delta N$ occurs for
all pairing symmetries. It would nevertheless be of interest to look for the helical to multiple-$q$ phase transition in Rashba superconductors. Fujimoto has pointed out that Fermi liquid
corrections for heavy fermion materials leads to a large
enhancement of a magnetoelectric effect that has the same origin
as the helical phase \cite{fuj05,kau05}. In our theory, this
enhancement is captured by an increase of $\delta N$.

\begin{figure}
\epsfxsize=3.5 in \center{\epsfbox{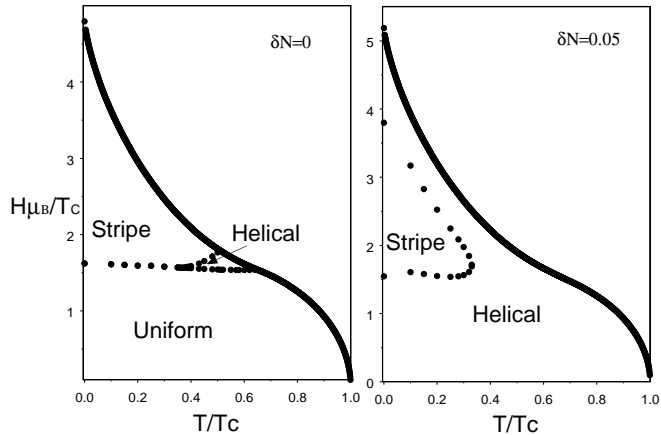}} \caption{Phase
diagram for an isotropic $s$-wave superconductor with a 3D
spherical Fermi surface when $\delta N=0$ and $\delta N=0.05$.
When $N_+$ and N$_-$ differ slightly, the helical phase dominates
the phase diagram.} \label{fig2}
\end{figure}

\subsection{Density of states in the helical phase}

\begin{figure}
\epsfxsize=3.5 in \center{\epsfbox{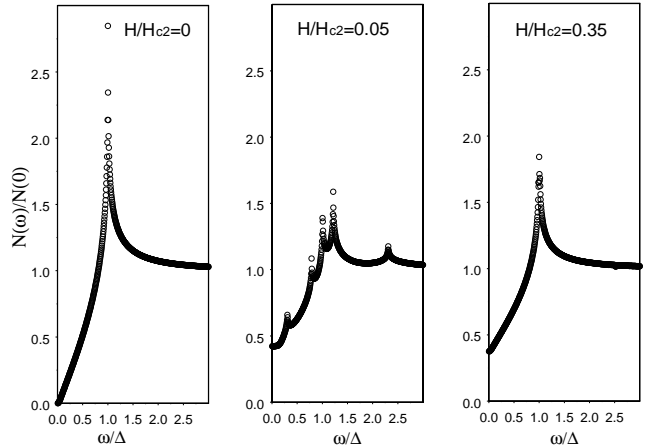}} \caption{Density
of states in the helical phases for increasing magnetic fields for
a $d$-wave superconductor. The high field density of states is field
independent and corresponds to that of an ungapped band and a
gapped band.} \label{fig3}
\end{figure}

\begin{figure}
\epsfxsize=3.5 in \center{\epsfbox{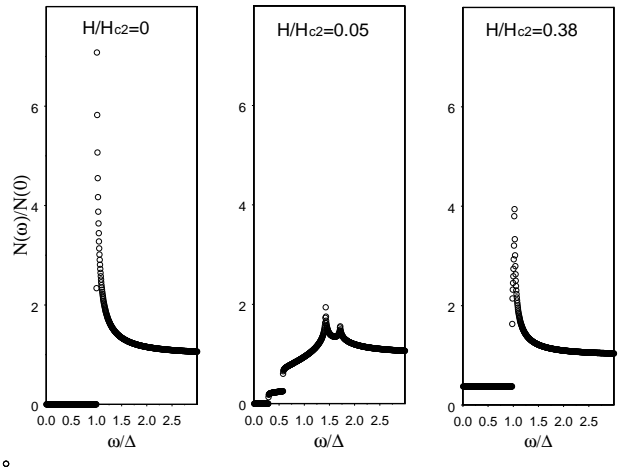}} \caption{Density
of states in the helical phases for increasing magnetic fields for
a $s$-wave superconductor. } \label{fig4}
\end{figure}

Having established the stability of the helical phase, it is of interest to determine
some of its physical properties. The density of states $N(\omega)$
is an important quantity for many properties and it is given by
\begin{equation}
N(\omega)=\sum_{j=\pm}N_j \hphantom{a}\Re \left\langle
\frac{|\tilde{\omega}_j|}{\sqrt{\tilde{\omega}_j^2-|\Delta\varphi_\Gamma(\hvk)|^2}}\right\rangle_{\hvk}
\end{equation}
where $\tilde{\omega}_{\pm}=\omega\pm\mu_B\hvg_{\hvk}\cdot \vH
+\vv_{\hvk}\cdot\vq$. Here we consider in detail $d_{x^2-y^2}$ pairing
symmetry on a 2D cylindrical Fermi surface at $T/T_c=0.15$ (and give analogous results for isotropic $s$-wave pairing). Note
that $d_{x^2-y^2}$ pairing is a possible pairing symmetry in
CePt$_3$Si, in which line nodes have been observed
\cite{bon05,iza05}. We set $\delta N =0.25$ and have checked that
the helical phase is the stable ground state.  For $T/T_c=0.15$,
$H_{c2}=21 H_P$ with the field applied along the gap maxima. The
evolution of $N(\omega)$  reveals two properties of interest that are revealed in Fig.~\ref{fig3}. The
most significant property is that for fields $H/H_{c2}\ge\approx
0.25$, $N(\omega)$ is field independent and has two contributions.
The first is a normal component that exists on one of the bands
and the second contribution is the usual $d$-wave density of
states from the second band. Similar behavior exists for isotropic
$s$-wave pairing as is shown in Fig.~\ref{fig4}. The observation that $N(\omega)$ for the one of
the bands corresponds to that of a normal metal is intriguing
because it exists even though both bands have the same pairing
amplitudes. Nevertheless, this result is intuitive because one
band prefers $v_F\vq=\mu_B \hat{z}\times\vH$ and is  not
frustrated while the other band prefers $v_F\vq=-\mu_B
\hat{z}\times\vH$ and therefore cannot pair the fermions that are
on the Fermi surface. This manifests itself as an increase of
$N(\omega)$. It is natural to expect that the frustrated band will
have its gap amplitude become zero at sufficiently large fields.
This possibility is not permitted by the theory considered here
because only spin-singlet pairing interactions have been included. Introducing an attractive spin-triplet pairing interaction will lead to a gap that vanishes on the frustrated band as the field is increased.
However, even if the frustrated band has zero gap amplitude, the
single particle density of states will remain qualitatively the
same.

The second property of interest occurs at low fields. In
particular, the single peak at $\omega=\Delta$ of the $H=0$
density of states splits into five peaks: one at $\omega=\Delta$
and the others at approximately $\omega=|\Delta\pm \mu_B H \pm v_F
q|$. This occurs for the field applied along the directions in which the gap is maximal. This property is anisotropic, for the field applied along the
gap nodes the $\omega=\Delta$ peak splits into four features that
occur at approximately $\omega=|\Delta\pm \mu_B H/\sqrt{2} \pm v_F
q/\sqrt{2}|$. This property may provide a means to probe the gap
symmetry in CePt$_3$Si. Note that if $\vq=0$, then an anisotropy
still exists in $N(\omega)$. At low fields, the $s$-wave density of states has
four features for all in-plane field orientations as is shown if Fig.~\ref{fig4}.

\section{Conclusions} We have derived the quasi-classical Eilenberger
theory describing Rashba superconductors. Using this theory, we
have examined the stability of helical and multiple-{\it q} phases
for Rashba superconductors in magnetic fields. We have found that
the helical phase is stable over a wide range of the phase
diagram. Finally, we have examined the single-particle density of
states in the helical phase and have revealed that at large fields
the it behaves qualitatively differently for the two spin-split
bands. On one band the density of states is completely normal,
while on the other band it remains gapped and resembles the
zero-field superconducting density of states.

We are grateful to P.A. Frigeri, N. Hayashi, and M.Sigrist for
many helpful discussions. This work was supported by the National
Science Foundation grant No. DMR-0381665.

\end{document}